# Expression profiles of TRPV1, TRPV4, TLR4 and ERK1/2 in the dorsal root ganglionic neurons of a cancer-induced neuropathy rat model

Ahmad Maqboul[1,2] and Bakheet Elsadek[2]

[1] Department of Anesthesiology and Operative Intensive Care Medicine, Charité Faculty of Medicine, Humboldt-Universität zu Berlin, Campus Mitte and Campus Virchow-Klinikum, Berlin, Germany
[2] Department of Biochemistry, College of Pharmacy, Al-Azhar University, Asyût, Egypt

## ABSTRACT

**Background:** The spread of tumors through neural routes is common in several types of cancer in which patients suffer from a moderate-to-severe neuropathy, neural damage and a distorted quality of life. Here we aim to examine the expression profiles of transient receptor potential vanilloid 1 (TRPV1) and of transient receptor potential vanilloid 4 (TRPV4), toll-like receptor 4 (TLR4) and extracellular signal-regulated kinase (ERK1/2), and to assess the possible therapeutic strategies through blockade of transient receptor potential (TRP) channels.

**Methods:** Cancer was induced within the sciatic nerves of male Copenhagen rats, and tissues from dorsal root ganglia (DRG) were collected and used for measurements of immunofluorescence and Western blotting. The TRPV1 antagonist capsazepine, the selective TRPV4 antagonist HC-067047 and the calcium ions inhibitor ruthenium red were used to treat thermal and/or mechanical hyperalgesia.

**Results:** Transient receptor potential vanilloid 1 showed a lower expression in DRGs on days 7 and 14. The expression of TRPV4, TLR4 and ERK1/2 showed an increase on day 3 then a decrease on days 7 and 14. TRPV1 and TLR4 as well as TRPV4 and ERK1/2 co-existed on the same neuronal cells. The neuropathic pain was reversed in dose-dependent manners by using the TRP antagonists and the calcium ions inhibitor.

**Conclusion:** The decreased expression of TRPV1 and TRPV4 is associated with high activation. The increased expression of TLR4 and ERK1/2 reveals earlier immune response and tumor progression, respectively, and their ultimate decrease is an indicator of nerve damage. We studied the possible role of TRPV1 and TRPV4 in transducing cancer-induced hyperalgesia. The possible treatment strategies of cancer-induced thermal and/or mechanical hyperalgesia using capsazepine, HC-067047 and ruthenium red are examined.

**Subjects** Molecular Biology, Neuroscience, Anesthesiology and Pain Management, Oncology, Pharmacology

# INTRODUCTION

In a recent study, we established a cancer-induced neuropathy model and determined the thresholds of cold allodynia and thermal and mechanical hyperalgesia. We have also confirmed the role of the transient receptor potential ankyrin 1 (TRPA1) in transducing cold allodynic pain (*Maqboul & Elsadek, 2017*). The objective of our current study is to understand the underlying molecular mechanisms of the transient receptor potential vanilloid 1 (TRPV1) and transient receptor potential vanilloid 4 (TRPV4), the innate immunity toll-like receptor 4 (TLR4) and the extracellular signal-regulated kinase (ERK1/2) through their expression profiles in the sensory neurons of this Copenhagen rat model. We also aim to assess a strategy for treatment of cancer-induced neuropathic pain using selective and non-selective transient receptor potential (TRP) antagonists and a calcium ions inhibitor.

The capsaicin receptor, TRPV1, functions as a thermal (*Samer, 2011*) or mechanical (*Cui et al., 2014*) transducer or both (*Walder et al., 2012*). Its expression increases in dorsal root ganglia (DRGs) upon injecting carcinoma cells of rat mammary gland in the bone marrow of rat's tibia (*Li et al., 2014*). In the same model of bone marrow, TRPV1 and lysophosphatidic acid (LPA) receptor 1 (LPA1) are both expressed in DRG neurons in which LPA potentiates TRPV1 current in DRGs (*Pan, Zhang & Zhao, 2010*). TRPV1 is co-expressed with TLR4 in enteric neurons of rat colon (*Filippova et al., 2015*). Moreover, TLR4 is expressed in trigeminal ganglia (TG) nociceptors (*Wadachi & Hargreaves, 2006*), and co-exists with TRPV1 on both rat TG and DRG neuronal cells (*Helley et al., 2015*). Therefore, it was important to investigate the potential of our model and the inoculated tumor cells to induce the expression of TLR4 and to initiate an innate immunity response.

Transient receptor potential vanilloid 4 is a polymodal ionotropic receptor activated by an array of agonists and stimuli. TRPV4 shows a spontaneous activity even without a specific activator's stimulation (*White et al., 2016*), and has, with other TRP channels, an important role through its expression in sensory neurons for mechanotransduction of pain associated with peripheral neuropathies (*Berrout et al., 2012*). ERK1/2 is a marker of cancer progress; it has a role in cell survival, growth regulation, proliferation, apoptosis and cell cycle's G1- to S-phase progression (*Lu & Xu, 2006*; *Meloche & Pouyssegur, 2007*). In our study, we concurrently studied the expression of ERK1/2 with TRPV4 to get an indication for tumor progression.

To conduct a treatment strategy, we hypothesized that these two receptors are continuously activated by the distant growing tumor in the afferent sensory neurons of cancer-induced Copenhagen rats. We have selected capsazepine, a TRPV1 antagonist and HC-067047, a selective TRPV4 antagonist for reversal of thermal and mechanical hyperalgesia respectively. They have a central amide group in common, capsazepine is a carbothioamide and HC-067047 a carboxamide, and an extensively explored parenteral routes. Before injection of these antagonists into the animals, we performed a docking study to find the possible conformation of capsazepine and HC-067047 in the binding pockets of their receptors and to determine the drug-receptor affinity. We also used ruthenium red, a calcium signaling inhibitor, due to its multiple mechanisms of action and its inhibition of intracellular calcium mobilization (*Eun et al., 2001*) to treat thermal as well as mechanical hyperalgesia.

# MATERIALS AND METHODS

## Cell line, antibodies and antagonists

Anaplastic tumor-1 (AT-1) cell line (Cat. #: 94101449) was purchased from ECACC, Salisbury, UK; anti-TRPV1 (goat, Cat. #: sc-12498) and anti-TLR4 (rabbit, Cat. #: sc-10741) antibodies from Santa Cruz Biotechnology, Inc., Heidelberg, Germany; anti-TRPV4 (sheep, Cat. #: ab63079 and rabbit, Cat. #: ab63003) antibodies from Abcam plc, Cambridge Science Park, Cambridge, UK; and anti-ERK1/2 (rabbit, Cat. #: 4695) antibody from Cell Signaling Technology Europe, B.V. Leiden, Netherlands. The secondary antibodies (H+L) Alexa Fluor® 594 (donkey anti-goat, Cat. #: A-11058) and Alexa Fluor® 488 (donkey anti-rabbit, Cat. #: A-21206) cross-adsorbed IgG were obtained from Invitrogen, Carlsbad, CA, USA, by ThermoFisher Scientific™, Inc., Rockford, IL, USA, Life Technologies GmbH, Darmstadt, Germany; and the Texas Red® AffiniPure IgG (donkey anti-sheep, Cat. #: 713-035-147) from The Jackson Laboratory "Jax®," Sulzfeld, Germany. Peroxidase-conjugated ((rabbit anti-goat, Cat. #: R1317HRP) and (goat anti-rabbit, Cat. #: 111-035-144)) IgG antibodies from Acris Antibodies GmbH, Herford, Germany and Jax® respectively.

2-Methyl-1-(3-morpholinopropyl)-5-phenyl-*N*-(3-(trifluoromethyl)phenyl)-1*H*-pyrrole-3-carboxamide (HC-067047; CAS: 883031-03-6; Cat. #: SML0143) and *N*-[2-(4-chlorophenyl)ethyl]-1,3,4,5-tetrahydro-7,8-dihydroxy-2*H*-2-benzazepine-2-carbothioamide (capsazepine; CAS: 138977-28-3; Cat. #: C191) were purchased from Sigma-Aldrich, Co., St. Louis, MO, USA, and the hexachlorotriruthenoxane tetradecaamine tetrahydrate (ruthenium red; CAS: 11103-72-3; Cat. #: R2751) from Sigma-Aldrich Chemie GmbH, Steinheim, Germany.

## Copenhagen rats (COP/CrCrl)

Inbred male Copenhagen rats were obtained from Charles River Laboratories International, Inc., Cologne, Germany. They arrived, through the Research Institutes for Experimental Medicine (FEM), Charité Faculty of Medicine, with initial weights about 250 g and were housed at standard conditions of number (not exceeding four and not less than two rats per cage), food and water supply (availability at all times) and light exposure (12 h light:12 h dark cycles). Rats were familiarized to human handling and the laboratory and the test enclosures from two to three weeks before the start of the experiments. Rats were euthanized if signs of pain or distress were observed. Euthanasia was performed in a separate room containing a fixed-size clear chamber and a $CO_2$ source (with a 10–30% displacement rate "1.8–5.3 L/min") (*AVMA Panel on Euthanasia, 2013*). The study was approved by the State Office for Health and Social Affairs (LAGeSo, Berlin, Germany) in adherence to the guidelines of the Charité Faculty of Medicine (Project Code: G 0314/13).

## Cancer cells inoculation and tissue collection

The AT-1 cells, derived from Dunning R3327 cell line, were cultivated in (RPMI 1640, L-glutamine, dexamethasone and 10% fetal bovine serum) medium and maintained at 37 °C and 5% $CO_2$ atmosphere. Copenhagen rats were anesthetized with isoflurane and then maintained at 2% isoflurane in $O_2$ inhalation throughout the operation time. The right hind leg was shaved and sterilized with alcohol and iodine. To expose the SN, an incision was made between the *gluteus superficialis* and the *biceps femoris* muscles (*Maqboul & Elsadek, 2017*). Cells were suspended in a pH-7.4 phosphate buffered saline (PBS) as a vehicle ($0.5 \times 10^6$ per 10 µL PBS) and slowly inoculated in the perineurium sheath of the SN using a 25 µL Hamilton® GASTIGHT® syringe 1702LT Series (Sigma-Aldrich Chemie GmbH, Steinheim, Germany). Sham-operated rats were injected with this PBS and used as a control in all the successive assays.

Metamizole sodium was injected as an analgesic and added to the drinking water for three days after surgery. Then, a daily checkup was conducted to ensure normal life activities like moving, walking and jumping, and to recognize any signs of pain or distress like curved posture, lack of grooming habits, decrease in eating or drinking or vocalization. Tissues of lumbar L3–5 DRG were collected in Eppendorf® Tubes, Eppendorf AG, Hamburg, Germany and immediately immersed in liquid nitrogen and stored at −80 °C until the time of experiment (*Maqboul & Elsadek, 2017*).

## Immunofluorescence double-staining

Perfusion and collection of DRG tissues ($n = 6$ rats) and sequential sectioning were performed according to our previous report (*Maqboul & Elsadek, 2017*). DRG sections were incubated overnight at room temperature with the primary antibodies. Then, they were washed and incubated again with the appropriate tagged secondary antibody (Alexa Fluor® 488 or 594 or Texas Red®). Nuclear DNA was labeled with 4′,6-diamidino-2-phenylindole. We used laser scanning microscope (LSM) (Zeiss® LSM 510; Carl Zeiss, Göttingen, Germany) to view the slides and to capture the images of TRPV1 with TLR4 (sham ($n = 17$) and 3 ($n = 21$), 7 ($n = 14$) and 14 ($n = 12$) days) and TRPV4 with ERK1/2 (sham ($n = 9$) and 3 ($n = 10$), 7 ($n = 12$) and 14 ($n = 10$) days).

## Software-aided counting

We referred to ImageJ 1.51n (Fiji) (*Schindelin et al., 2012*) as a tool for software-aided counting due to its accuracy and lack of vagaries and bias (*Guillery & August, 2002*). The intensities of regions of interest (ROIs) were measured to get the relative values of the stained neuronal cells. To measure the pixel intensity values (PIVs) of these ROIs in the whole image (according to ImageJ/Fiji User Guide 1.46r), photos were converted from a LSM to a tagged image file (TIF) format. The software was adjusted to auto-scale the 8-bit images, binaries were generated, and the intensity values of backgrounds were subtracted (examples are shown in Fig. 1). Each ROI was measured, and the results of each image were exported to a specific comma-separated values (CSV) file.

## Gel electrophoresis and Western blotting

TRPV1 and TRPV4 were semi-quantified in DRGs using Western blotting. Proteins were extracted and measured using Pierce™ BCA Protein Assay Kit (ThermoFisher Scientific™ Inc., Waltham, MA, USA). The separation of TRPV1 and TRPV4 was done on 8% sodium dodecyl sulfate (SDS)-polyacrylamide gel and TGX Stain-Free Gel (4–15%) using 15 and 10 μg protein per lane respectively. They were then blotted on nitrocellulose membrane using the Trans-Blot® Turbo™ Transfer System (Bio-Rad, Hercules, CA, USA) and incubated overnight at 4 °C with the corresponding primary antibody (goat anti-TRPV1; 1:200 and rabbit anti-TRPV4; 1:500). We incubated the blots again for 120 min at room temperature with the corresponding horseradish peroxidase (HRP)-conjugated secondary antibody (anti-goat; 1:10,000 and anti-rabbit; 1:20,000).

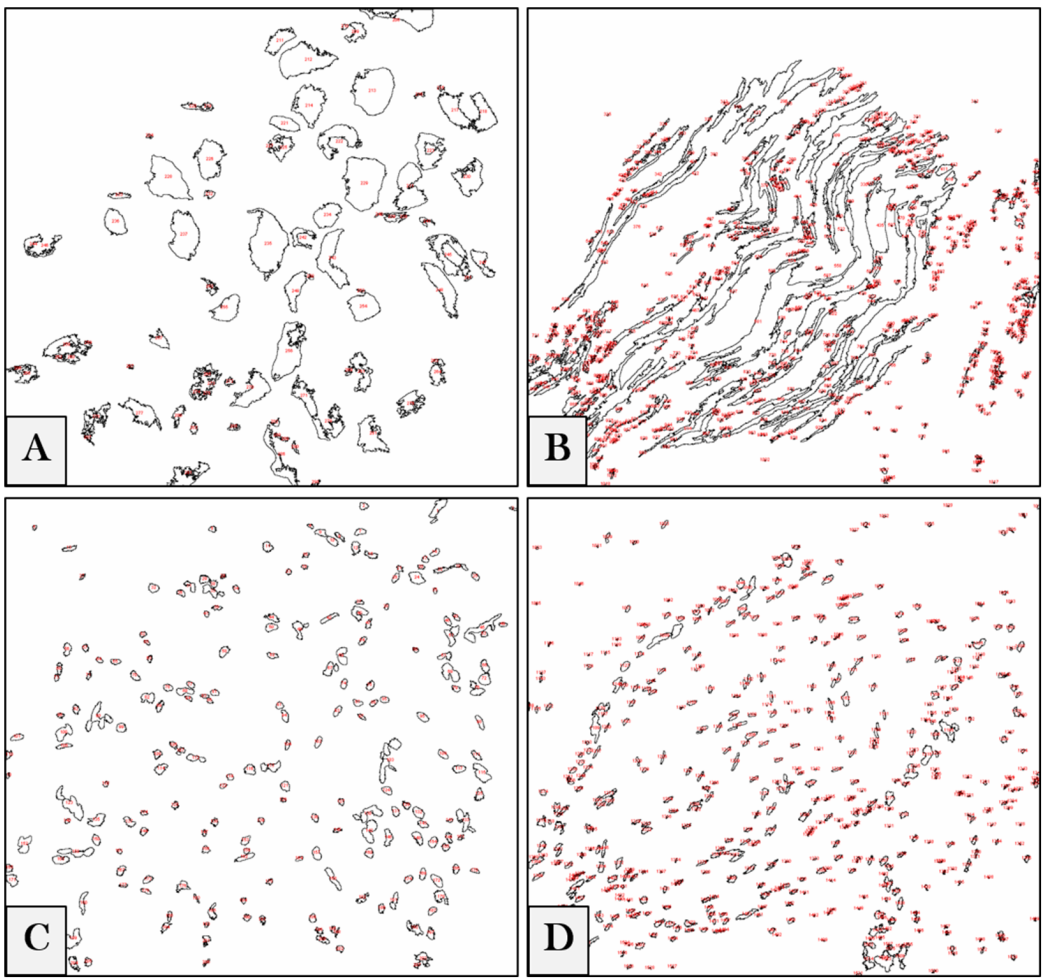

Visualization of bands was performed using the ECL Kit (Amersham Pharmacia Biotech, Piscataway, NJ, USA) and images were captured by ChemiDoc™ MP Imaging System (Bio-Rad, Hercules, CA, USA). The monoclonal anti-β-actin-peroxidase (A3854; Sigma-Aldrich, Co., St. Louis, MO, USA"; 1:25,000) was used to view the housekeeping protein band (*Towbin, Staehelin & Gordon, 1992*). The images' brightness and contrast were adjusted using Fiji software, and the backgrounds were subtracted (*Schindelin et al., 2012*). To measure the integrated densities, ROIs were selected for all bands ($n = 10$ samples (five blots for each protein in duplicates)) and calculated as percentages to the corresponding sham.

## Molecular docking

Crystal structures of TRPV1 and TRPV4 were retrieved from protein data bank (PDB codes: 5RIX and 3W9F, respectively) (*Berman, Henrick & Nakamura, 2003*) and loaded to Molegro Virtual Docker 2013.6.0.0 (CLC bio; QIAGEN Bioinformatics, Aarhus, Denmark) (*Thomsen & Christensen, 2006*). Protein structures were prepared and refined, to ensure chemical correctness, the non-bonded oxygen atoms of water were removed, and the missing hydrogens were added. Formal charges and potential steric clashes via protein minimization were assigned. ChemBio3D Ultra 16 (CambridgeSoft Corp., Cambridge Scientific Computing, Inc., Richmond, CA, USA) was used to draw the 3D structures of capsazepine and HC-067047. The TRP antagonists were further pre-optimized using MarvinSketch v6.1.0 (ChemAxon Ltd., Budapest, Hungary) with MM force field and saved in a Tripos Mol2 (.mol2) file format.

Optimization of conformations were converged to a gradient root-mean-square deviation (RMSD) below 0.05 kJ/mol or continued to a maximum of 1,000 iterations at which point there were negligible changes in RMSD gradients. The binding sites were defined by constraints at XYZ 117.68, 137.93, 128.79 and −9.98, 12.24, 20.81 for TRPV1 and TRPV4 respectively. The search began with a rough positioning and scoring phase that significantly narrowed the search space and reduced the number of poses to be further considered to a few hundreds. The selected poses were minimized and then the 5–10 of lowest-energy, obtained in this fashion, were selected.

## Treatment strategies

The animal responses to thermal and mechanical stimuli were assessed by Hargreaves (*Hargreaves et al., 1988*) and von Frey (*Dixon, 1965*) methods using Ugo Basile® Plantar Test and Touch Test® Sensory Evaluators (North Coast Medical, Inc., Morgan Hill) respectively. Rats were placed on a glass or a wire mesh table (for Hargreaves or von Frey test respectively) and restrained under transparent plastic ventilated boxes 30 min before the test for acclimation purposes. To prevent tissue damage from increased temperatures, the IR source was adjusted to a 20 s cut-off value. Heat withdrawal latency and force threshold were measured as a mean of duplicate and single measurements respectively in sham and cancer-inoculated animals. While the latencies were measured upon the elevation of the rat paw and the IR source was automatically switched-off, the determination of the withdrawal force was measured by reaching a response upon moving up-and-down between filaments of different strengths. Both tests were randomized, and the examiner was blinded for the animal assignment to different sham or cancer groups.

## Antagonizing TRPV1 by capsazepine

Capsazepine was solubilized and injected at the time of the experiment. The concentrations of the subcutaneous (S.C.) doses were 1, 5 and 10 mg/kg animal body weight ($n = 6$ rats for each concentration). Hargreaves' plantar test was applied to measure the thermal hypersensitivity evoked by anaplastic tumor's invasion of the sciatic nerves. Responses were recorded in duplicates on five pre-scheduled points (30, 40, 50, 70 and 90 min after six days of tumor inoculation).

## Selectively blocking TRPV4 by HC-067047

HC-067047 was dissolved in dimethyl sulfoxide and diluted with ethyl alcohol (1:9 v/v) and then with water (1:9 v/v). Doses used for injection were 10 ($n = 9$ rats) and 20 ($n = 7$ rats) mg/kg body weight. Von Frey test was carried out on day 7 after AT-1 cells injection. Responses were acquired in single measurements on five timepoints (30, 40, 50, 70 and 90 min) after HC-067047 S.C. administration.

## Inhibiting Ca²⁺ ions by ruthenium red

The calcium ($Ca^{2+}$) ions inhibitor ruthenium red was S.C. injected on day 7 to study its effects for treatment of cancer-induced thermal ($n = 7$ rats for each concentration) and mechanical (0.5 ($n = 7$ rats), 1 ($n = 10$ rats) and 2 ($n = 7$ rats) mg/kg body weight) hyperalgesia. Thermal and mechanical sensations were estimated by Hargreaves and von Frey tests respectively. Measurements were recorded on 15, 30, 60 and 120 min after ruthenium red injection.

## Data analysis and manuscript writing

GraphPad Prism® (GraphPad Software, Inc., San Diego, CA, USA) and Microsoft® Excel 2016 MSO (16.0.8229.2086) (Microsoft® Deutschland GmbH, Berlin, Germany) were used for doing statistics, graphs and tabular data descriptions. Data were expressed as mean ± standard error of mean (SEM). Results were tested for possible outliers. One-way analysis of variance (ANOVA) followed by Dunnett's post hoc test were used to make pairwise or versus-control comparisons. Kruskal–Wallis test was used to analyze variances on ranks if normality test failed. Values of at least $P < 0.05$ were considered statistically significant. Microsoft® Office 365 ProPlus (Word) was used for typing texts, and EndNote™ X7.5 (Thomson Reuters Deutschland GmbH, Frankfurt, Germany) for managing citations.

# RESULTS

The two baselines of thermal and mechanical hyperalgesia were previously measured from day 0 to day 14 in both sham-operated and cancer AT-1 cells-injected rats (*Maqboul & Elsadek, 2017*). In cancer animals, the sensitivity to thermal stimuli decreased on day 2 and did not change on day 4 but increased on days 6–14. On the other hand, a significant increase in the degree of mechanical sensation was observed from day 6 to day 14.

To study the molecular mechanisms underlying the process of tumor infiltration of the peripheral nerves, we quantified two key transducers, TRPV1 and TRPV4, within the neuronal cells by immunofluorescence and Western blotting.

## TRPV1 and TLR4

Fluorescence images of the double-stained capsaicin (TRPV1) and the lipopolysaccharide toll-like (TLR4) receptors showed no changes in TRPV1 positive immunoreactive (+IR) cells on day 3 then decreased on days 7 and 14. On the other hand, TLR4 +IR cells were highly increased on day 3 then also decreased on days 7 and 14. The images also verified that both TRPV1 and TLR4 were co-expressed on the same cells (Fig. 2).

Measurements of PIVs of +IR cells were performed to each stain alone and comparisons were made against sham group. TRPV1 images showed no changes on day 3 (98.3 ± 1.97%), and then significantly decreased on days 7 (61.7 ± 1.15%) and 14 (76.6 ± 1.95%) as compared to sham (100 ± 2.46%). PIVs of TLR4 +IR cells increased on day 3 (128.2 ± 2.51%) then substantially decreased on days 7 (80.9 ± 1.97%) and 14 (84.6 ± 2.37%) as compared to (100 ± 2.02%) of sham. Density values of nuclei gradually decreased on days 3 (96.9 ± 0.95%), 7 (46.0 ± 0.64%) and 14 (49.7 ± 0.80%) as compared to (100.0 ± 1.19%) of sham (Fig. 3).

As we expected, the results of TRPV1 blotting revealed an electrophoretic profile similar to that of immunofluorescence. Lower integrated density was measured on days 7 (69 ± 3.1%) and 14 (56 ± 5.5%) as compared to (97 ± 4.3%) of sham (Fig. 4; the whole blots are shown in Supplemental File).

## TRPV4 and ERK1/2

Immunofluorescent photos showed a significant increase in the number of TRPV4 and ERK1/2 +IR neurons on day 3 and then decreased on days 7 and 14. Photos also confirmed the co-expression of TRPV4 and ERK1/2 on the same neuronal cells (Fig. 5).

The expression of TRPV4 and ERK1/2 was confirmed by measuring the PIVs of +IR cells of the two proteins as compared to the sham group. The expression of TRPV4 +IR cells increased on day 3 (108.9 ± 1.49%) then decreased on days 7 and 14 with (54.7 ± 0.67%) and (66.5 ± 0.81%) respectively as compared to sham (100 ± 1.47%). PIVs of ERK1/2 +IR cells were increased on day 3 (136.5 ± 2.34%) then decreased on day 7 (78.0 ± 1.15%) and increased again on day 14 (107.1 ± 1.6%) as compared to sham (100.0 ± 1.56%). Density values of nuclei did not change on day 3 (103.2 ± 1.61%) and then decreased on days 7 (65.1 ± 0.93%) and 14 (78.9 ± 1.18%) as compared to (100.0 ± 1.84%) of sham (Fig. 6).

The mean integrated densities of TRPV4 blotting bands showed a significant increase on day 3 (141 ± 9.7%) followed by a decrease on days 7 (57 ± 3.8%) and 14 (40 ± 5.8%) as compared to (101 ± 0.65%) of sham (Fig. 7; the whole blots are shown in Supplemental File).

## Molecular docking

The minimum energy of interaction was represented by different scoring functions using MolDock score. Capsazepine interacted with the active site's amino acids of TRPV1 with three hydrogen bonds between capsazepine NH and OH groups with Thr550, Ala546 and Phe543 residues with bond lengths 2.61, 3.08 and 2.73 Å, respectively. In addition to hydrogen bonds, there were hydrophobic interaction between dihydroxyphenyl moiety and Phe543 and Phe591 residues, and chlorophenyl ring and Tyr511 and Tyr554 amino acids (Fig. 8A).

# PeerJ

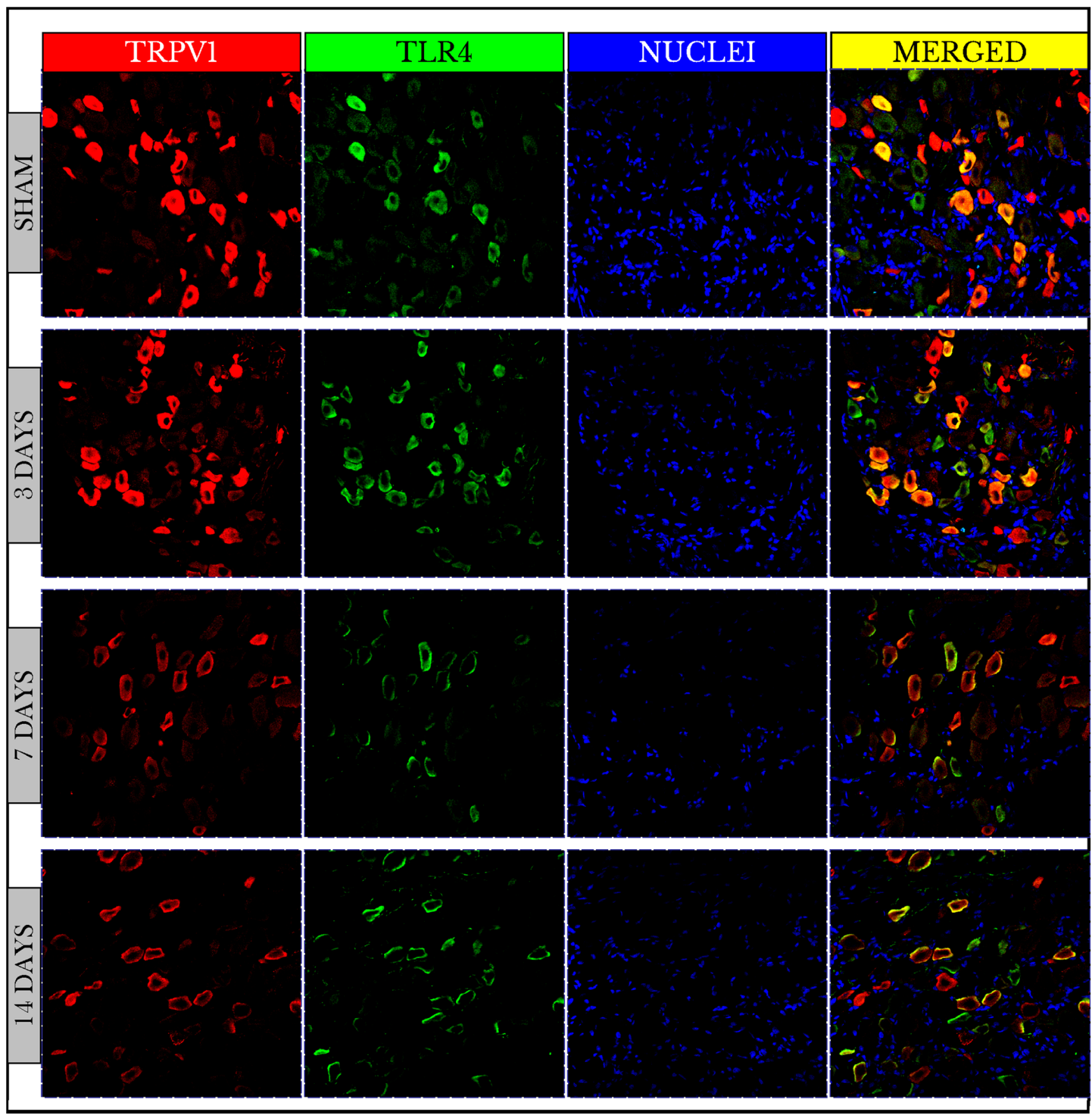

**Figure 2  Fluorescence micrographs of DRG neurons in sham and cancer animals (after 3, 7 and 14 days of injection).** TRPV1 is stained by a *red* fluorescent dye, TLR4 by *green*, and nuclei by *blue*. The expression of TRPV1 and TLR4 takes place on the same cells (*orange/yellow*). The number of TRPV1 +IR cells decreases on days 7 and 14. The number of TLR4 +IR cells increases on day 3 and then decreases on days 7 and 14. The number of nuclei gradually decreases over days 3–14.

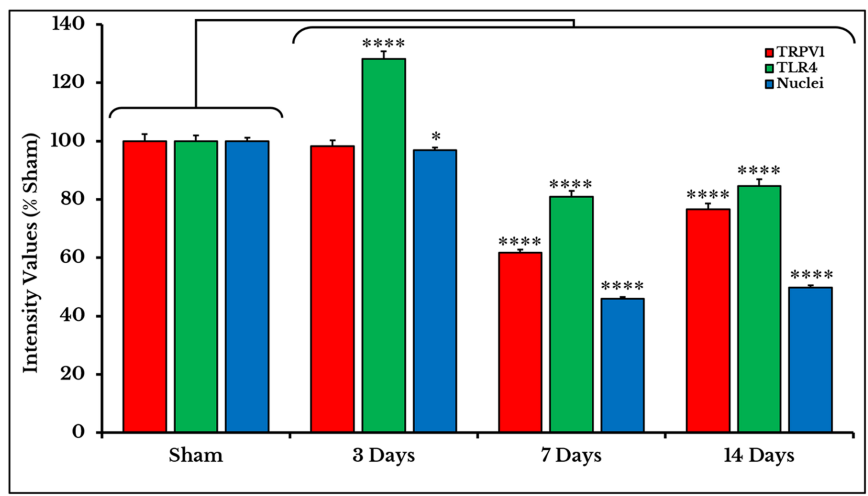

**Figure 3** **Quantifying PIVs of TRPV1 and TLR4 +IR cells in DRGs of sham and cancer animals (after 3, 7 and 14 days of injection).** TRPV1 +IR cells (*red*) show a significant decrease on days 7 and 14 ($P < 0.0001$). TLR4 +IR cells (*green*) show an increase on day 3 ($P < 0.0001$) followed by a decrease on days 7 and 14 ($P < 0.0001$). Nuclei (*blue*) show a decrease ($P < 0.05$) on day 3, then on days 7 and 14 ($P < 0.0001$). Full-size ☑ DOI: 10.7717/peerj.4622/fig-3

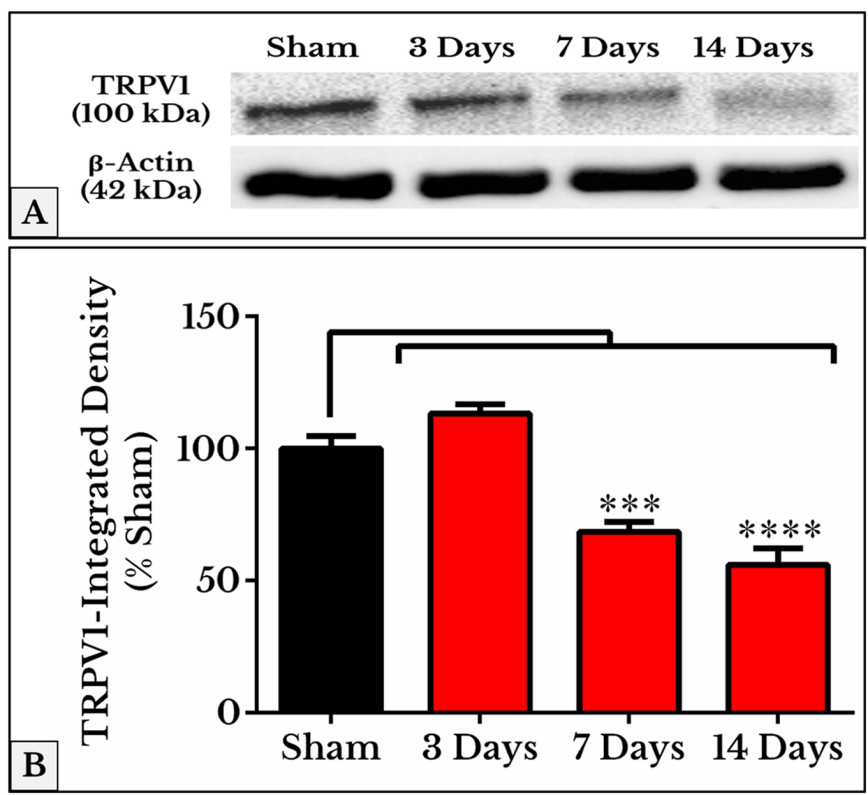

**Figure 4** **Representative immunoblots and quantitative densitometry of TRPV1 in DRGs of sham and cancer animals (after 3, 7 and 14 days of injection).** (A) The bands show TRPV1 and the internal standard, β-Actin, blotted. (B) TRPV1 band shows a decrease in integrated density on days 7 ($P < 0.001$) and 14 ($P < 0.0001$). Full-size ☑ DOI: 10.7717/peerj.4622/fig-4

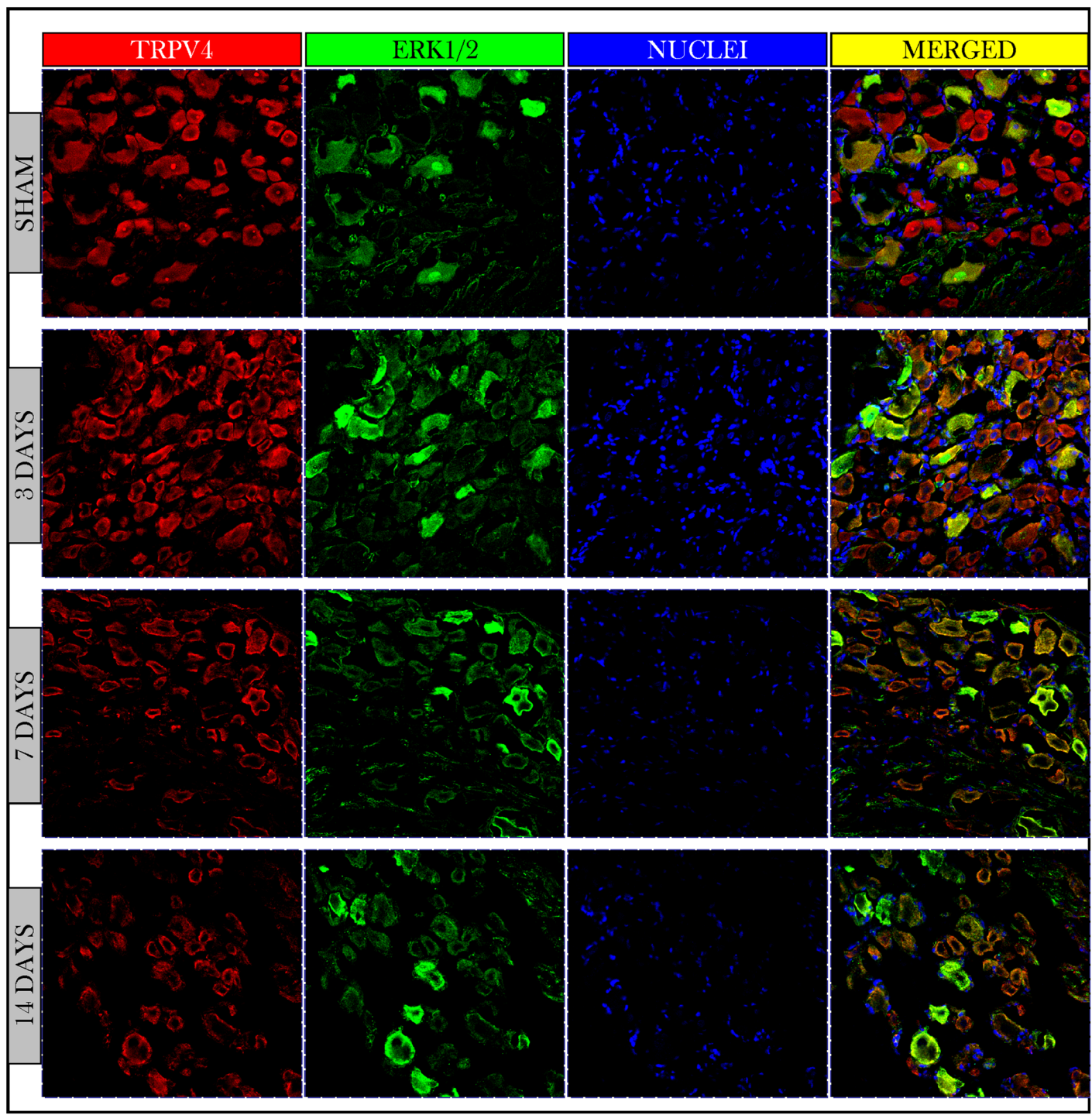

**Figure 5  Immunofluorescent micrographs of DRG neurons in sham and cancer animals (after 3, 7 and 14 days of injection).** TRPV4 is stained by a *red* fluorescent dye, ERK1/2 by *green*, and nuclei by *blue*. TRPV4 and ERK1/2 are co-expressed on the same cells (*orange/yellow*). The number of TRPV4 and ERK1/2 +IR cells shows an increase on day 3 followed by a decrease on days 7 and 14 as compared to sham.

Full-size  DOI: 10.7717/peerj.4622/fig-5

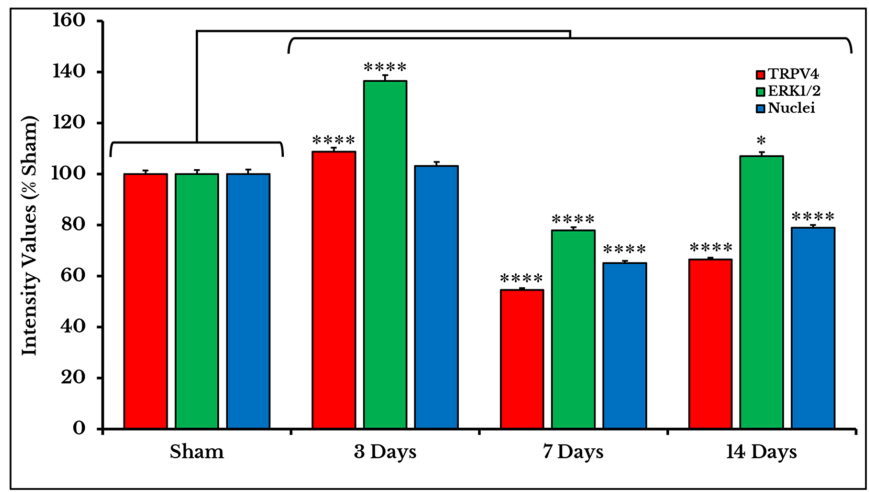

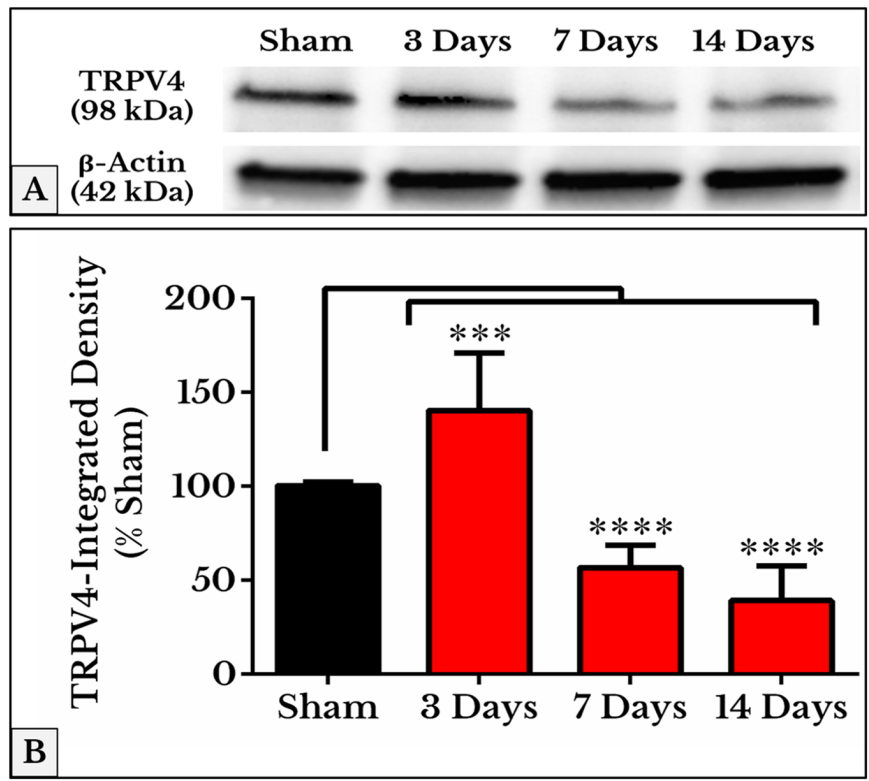

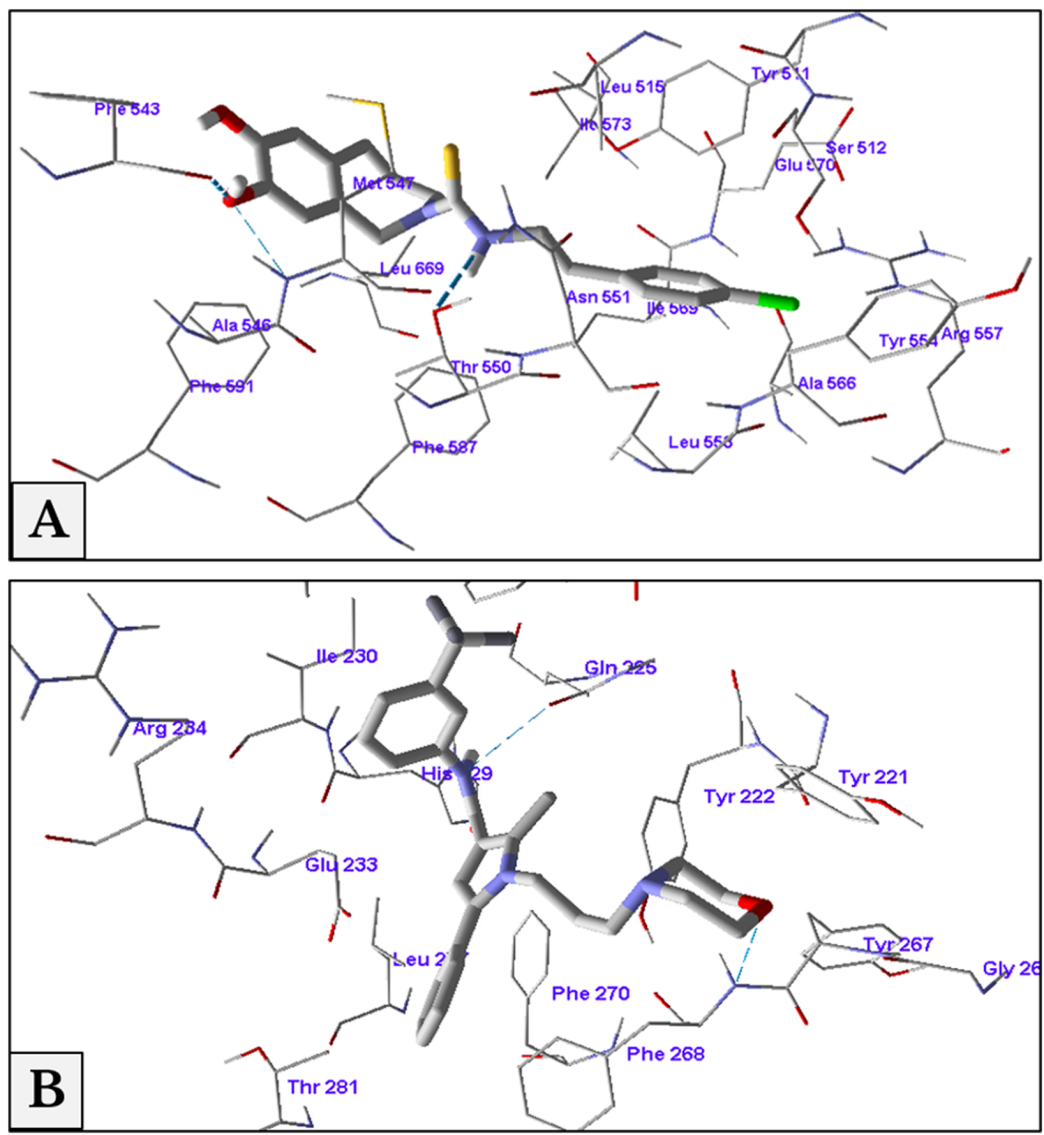

**Figure 8 Antagonist-receptor docking modes.** Binding of capsazepine (A) and HC-067047 (B) into their active sites of TRPV1 and TRPV4 receptors respectively.

Full-size ⬜ DOI: 10.7717/peerj.4622/fig-8

Docking of HC-067047 into the active site of TRPV4 structure gave a MolDock score of −134.1 indicating a good affinity of the antagonist with the receptor. HC-067047 made two hydrogen bonds with the active site's amino acids. These were O atom of morpholine with NH group of Phe268 and NH group with CO group of Gln225 with bond lengths 2.64 and 2.74 Å respectively. The phenyl ring of HC-067047 made also a hydrophobic interaction with Phe268 and Phe270 (Fig. 8B).

## Capsazepine and thermotransduction

Subcutaneous injection of the vanilloid receptor antagonist, capsazepine, at doses of 1, 5 and 10 mg/kg rat body weight transiently reversed the cancer-induced thermal hyperalgesia as compared to the previously measured thermal pain on day 6. The vehicle group did not show any change after S.C. injection of capsazepine. For the 1 mg/kg dose, the latency (time required to give a response) significantly increased at minutes 30 and 40 with mean and SEM of ($11 \pm 0.89$) and ($9.2 \pm 0.76$) respectively. For the 5 mg/kg dose, the latency increased at minutes 30 ($11 \pm 0.95$), 40 ($11 \pm 1.1$) and 50 ($8.2 \pm 0.66$). For the 10 mg/kg dose, it increased at minutes 30 ($12 \pm 0.8$), 40 ($11 \pm 1.0$), 50 ($9.3 \pm 0.98$) and 70 ($8.6 \pm 0.95$) (Fig. 9).

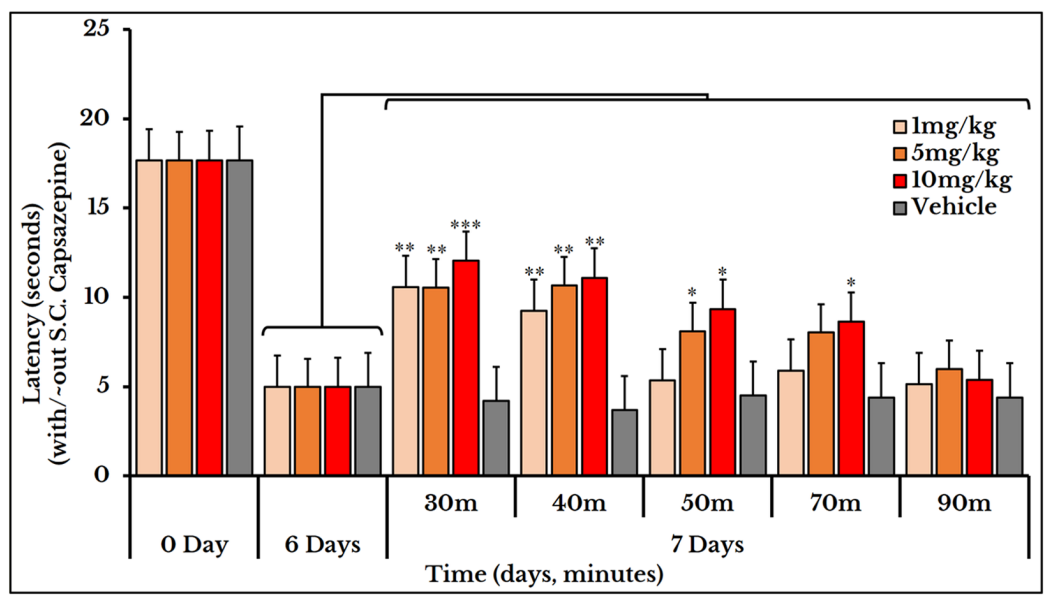

## HC-067047 and mechanotransduction

The vehicle group did not show any change after S.C. injection of HC-067047. For the 10 mg/kg dose, the 50%g threshold significantly increased at minutes 30 and 40 with means ($5.1 \pm 0.59$) and ($3.8 \pm 0.78$) respectively as compared to day 6 ($0.82 \pm 0.15$). Then it relapsed again at minutes 50 ($1.6 \pm 0.41$), 70 ($0.66 \pm 0.07$) and 90 ($0.61 \pm 0.09$) as compared to ($2.4 \pm 0.41$) of day 6. The 20 mg/kg dose showed a higher and longer blocking effect. The 50%g threshold (represented by tolerance) substantially increased at minutes 30 ($11 \pm 0.91$), 40 ($15 \pm 1.5$) and 50 ($10 \pm 1.2$) as compared to day 6 ($2.4 \pm 0.41$). Then it decreased back at minutes 70 ($4.7 \pm 0.97$) and 90 ($2.0 \pm 0.26$) as shown in Fig. 10.

## Ruthenium red and thermo- and mechanotransduction

Ruthenium red was used for treating both thermal and mechanical hyperalgesia. Three doses of ruthenium red (0.5, 1 and 2 mg/kg rat body weight) were used to measure the thermal hyperalgesia. For the 0.5 mg/kg dose, the thermal pain was decreased (i.e., an increased latency) at minutes 15 ($17 \pm 0.85$), 30 ($13 \pm 0.79$) and 60 ($11 \pm 0.58$) as compared to ($5.7 \pm 0.68$) on day 6. For the 1 mg/kg dose, the latency increased at minutes

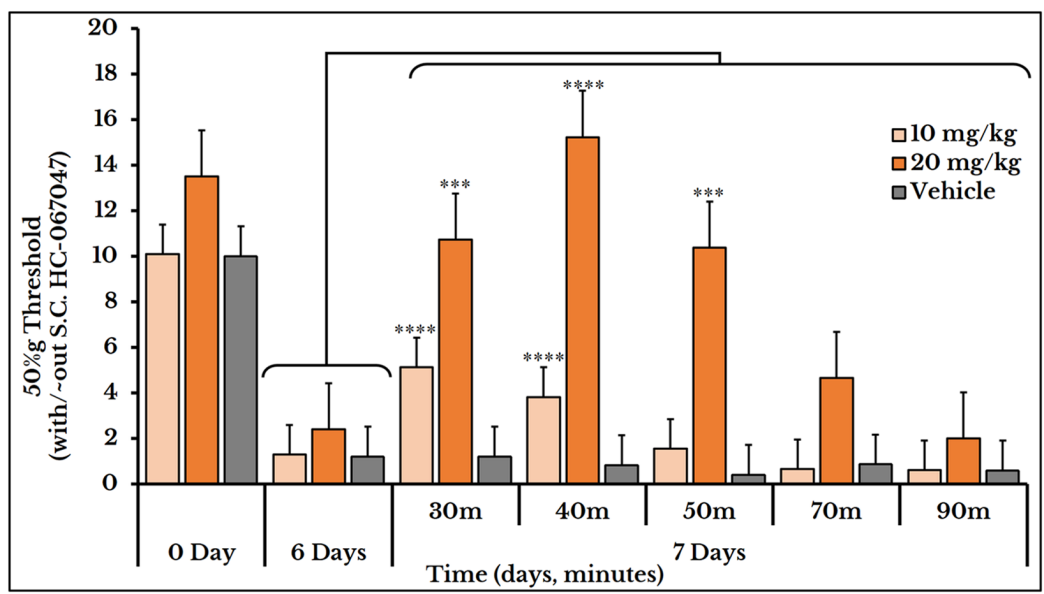

15 (18 ± 0.95), 30 (14 ± 0.69) and 60 (15 ± 1.1) as compared to (8.5 ± 1.3) on day 6. For dose 2 mg/kg, it increased at minutes 15 (18 ± 0.80), 30 (17 ± 0.73) and 60 (16 ± 0.94) as compared to (5.7 ± 0.68) on day 6 (Fig. 11).

We used the same three doses (0.5, 1 and 2 mg/kg body weight) for mechanical sensitivity assessment. For the vehicle-injected animals, the 50% g threshold values did not change on day 7 (all timepoints) and also for the two doses of 0.5 and 1 mg/kg. For the 2 mg/kg dose, it increased at minute 15 (9.2 ± 1.5) and 30 (11 ± 0.91) and then decreased again at minutes 60 (31 ± 0.5) and 120 (1.8 ± 0.15) as compared to (2.5 ± 0.62) of day 6 (Fig. 12).

## DISCUSSION

Based on our model of cancer-induced neuropathy, we have investigated the molecular mechanisms underlying the process of pain generation. One of the important characteristics of our model is that the thresholds of thermal and mechanical hyperalgesia were reached on day 6. This early neural invasion is due to the immediate contact with the nerve and the earlier induction of peripheral neuropathy, which was reported in other studies after two or three weeks (Hald et al., 2009; Shimoyama et al., 2002). In addition, the nature of the AT-1 cell line, of being non-metastasizing, androgen independent and very fast growing facilitated its transplantation into male Copenhagen rats (major histocompatibility complex haplotype RT1^{av1}) as well as the production of an efficient, fast and reproducible cancer model. Rats were prepared for behavioral experiments within only six days after inoculation of cancer cells, and tissues were collected after three, seven or 14 days.

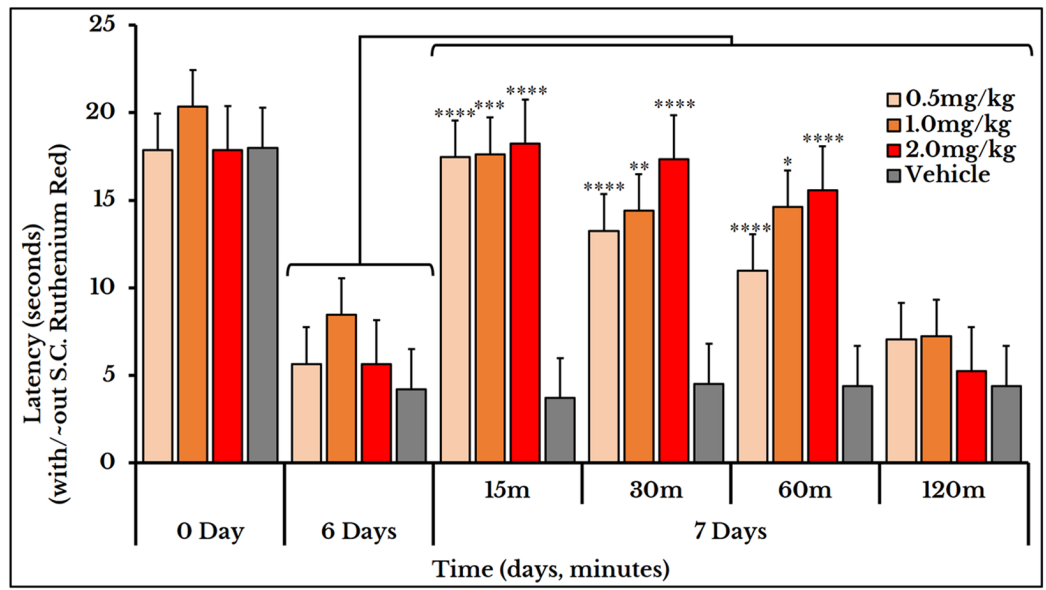

**Figure 11 Cancer-induced thermal hyperalgesia and its treatment using the calcium ions inhibitor ruthenium red.** Ruthenium red is S.C. injected at doses of 0.5, 1 and 2 mg/kg. Thermal pain is significantly reversed, using the three doses, at 15, 30 and 60 min with varied *P* values.

Full-size ☒ DOI: 10.7717/peerj.4622/fig-11

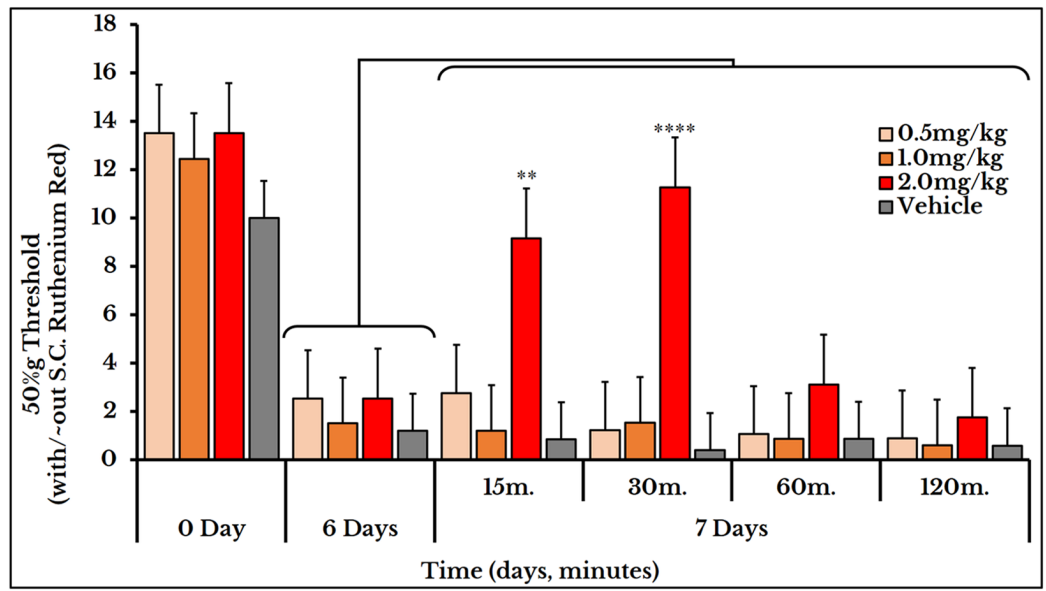

**Figure 12 Cancer-induced mechanical hyperalgesia and its treatment using the calcium ions inhibitor ruthenium red.** Ruthenium red is S.C. injected at doses of 0.5, 1 and 2 mg/kg. Pain is significantly treated using the 2.0 mg/kg dose at minute 15 ($P < 0.01$) and 30 ($P < 0.0001$).

Full-size ☒ DOI: 10.7717/peerj.4622/fig-12

The development of such a local non-metastasizing malignancy showed major molecular changes in the afferent DRG sensory neurons. Here, we have studied the expression of the TRP channels TRPV1 and TRPV4 responsible for thermal and mechanical hyperalgesia respectively (*Sousa-Valente et al., 2014*). In addition to TRPs, we have examined the expression of TLR4 and the ERK1/2 in DRG neurons.

## Expression of TRPV1 and TLR4

The significant decrease of TRPV1 expression is consistent with a similar investigation on papillary urothelial carcinoma (*Sterle, Zupancic & Romih, 2014*). However, some researchers reported its increased (*Asai et al., 2005*; *Niiyama et al., 2007*; *Zhang et al., 2012*) or unaltered expression (*Nagae, Hiraga & Yoneda, 2007*). Here, we also confirm the relationship between TRPV1 and the gram-negative bacterial receptor TLR4 by their concurrent expression on DRG cells of the cancer-induced rat model (*Lin et al., 2015*). The earlier elevation of TLR4 +IR cells on day 3 followed by their decline is indicative for an immediate immune response and a late neural damage. Some researchers reported the dependence of TLR4 expression on TRPV1 (*Hakimizadeh et al., 2017*) while others described the activation of TRPV1 by LPS (*Ferraz et al., 2011*). This led to a proposed intracellular signalling via the activation of TRPV1 by TLR4 (*Assas, Miyan & Pennock, 2014*).

## Expression of TRPV4 and ERK1/2

The early increased expression of ERK1/2 in DRGs on day 3 is in agreement with other studies reporting its phosphorylation and hyperactivity in certain cancer cell lines (*Park et al., 2010*; *Steinmetz et al., 2004*) and chemotherapy-induced neuropathy (*Reyes-Gibby et al., 2015*) and the persistent pain-induction. This increased expression is an indicator for the progression of cancer cells. Its eventual decrease after day 3 could be explained by the ultimate nerve damage and neuronal death.

The concurrent expression of TRPV1 with TLR4 (*Filippova et al., 2015*; *Helley et al., 2015*; *Li et al., 2015*) and TRPV4 with ERK1/2 (*Barabas, Kossyreva & Stucky, 2012*; *Nakatsuka & Iwai, 2009*) on the same neurons implies mutual synergistic functions of these proteins in pain transduction. The immunofluorescent images of +IR cells comply with a similar trend of gel electrophoresis and Western blotting. This similarity is significant in both TRPV1 and TRPV4 blots. Levels of the housekeeping β-actin did not show any alteration between sham and all cancer treated groups in agreement with other studies on other types of cancer (*Bandyopadhyay et al., 2004*; *Wei et al., 2017*).

## Decreased expression but high activation

The decreased expression of TRPV1 and TRPV4 in cancer-bearing animals, as compared to sham, does not necessitate their decreased activity. It may be rather associated with their continuous activation by the distant invasive tumor and the subsequently observed increase in nociception. This conclusion is strengthened by these two studies: First, the low expression of stimulator of interferon genes (STING) was positively correlated to signs of tumor invasion (*Song et al., 2017*); and second, UMUC14 cells with low mitogen-activated protein kinase phosphatase 1 (MKP-1) expression were more invasive than UMUC6 cells with high MKP-1 (*Shimada et al., 2007*).

## Reversal of thermal and mechanical hyperalgesia

With the continuous growth of cancer cells, pressure was gradually exerted on the injection site of the sciatic nerve. This pressure led to an increase in the hypersensitivity to thermal and mechanical stimuli and produced a spontaneous pain behavior. Molecular docking studies confirmed the same docking pattern of capsazepine previously reported (*Gao et al., 2016*) and represented a new mode for HC-067047 binding. Both antagonists, capsazepine and HC-067047, showed good binding affinities to their receptors, TRPV1 and TRPV4, respectively.

Capsazepine was efficiently used for reversal of thermal hyperalgesia which is probably due to its blockade of TRPV1 receptor. This blockade is supported by similar reports on other models of squamous cell carcinoma (*Asai et al., 2005*; *Shinoda et al., 2008*), osteosarcoma (*Menendez et al., 2006*) and mammary carcinoma (*Naziroglu et al., 2017*) for reversing the cancer-produced thermal and/or mechanical hyperalgesia. In a previous study, capsazepine was reported to be effective in reversing capsaicin-produced thermal hyperalgesia in rats only when injected before capsaicin (*Walker et al., 2003*). This can be explained by the structural similarity between capsaicin and capsazepine; both has the common amide group (the former is a nonenamide and the latter is a carbothioamide) and a peripheral substituted phenyl ring (a hydroxy-methoxyphenyl in the agonist and a dihydroxyphenyl "of benzazepine" in the antagonist). This means that the binding affinity of capsazepine exceeds that of capsaicin (*Wahl et al., 2001*) (at concentrations $\geq$ 30 mg/kg S.C.), and once being embarked upon its target receptor, it will compete as long as its concentration is high.

HC-067047, the selective TRPV4 antagonist, showed a reversal of cancer-induced mechanical hyperalgesia in rats. This pyrrole-derivative (HC-067047) was introduced as an effective drug for the treatment or relapse of several neuropathies (e.g., alleviation of mechanical and thermal hyperalgesia in alcohol- and high fat (AHF)-induced pancreatitis rat model (*Zhang et al., 2015*), and the inhibition of mechanical hyperalgesia in bradykinin-induced mice model (*Costa et al., 2017*)). In our cancer-neuropathy model, pain was induced on day 6 compared to three weeks in the AHF-induced pancreatitis rat model, and a 5 or a 10 mg/kg dose was sufficient to pain reversal persisting for ~3 h in the AHF (*Zhang et al., 2015*) or 4 h in the chemotherapy-induced (*Costa et al., 2017*) model compared to 50 min in our model. This may be explained by the severity of the cancer cells compared to these two inflammatory models and the earlier immune response to tumor invasion. It may also highlight the differences between the intraperitoneal route in these two models and the S.C. route in ours which may be the result of differential drug distribution.

Ruthenium red was used as $Ca^{2+}$ ions inhibitor for treatment of both thermal and mechanical hyperalgesia. This dual action is consistent with other studies showing its role for treatment of thermal hyperalgesia in a paclitaxel-induced model (*Hara et al., 2013*), mechanical allodynia in a diabetes-induced model (*Cui et al., 2014*) and both thermal and mechanical hyperalgesia in a squamous cell carcinoma model (*Shinoda et al., 2008*). By comparing the inhibitory effect of ruthenium red, we found that it had a higher and a long-lasting efficacy for treating thermal than mechanical hyperalgesia suggesting a preferential TRPV1 over TRPV4 antagonism. This effect may be attributed to its multiple mechanisms of action and different pathways and target receptors other than the TRP channels.

Increasing the dose of capsazepine from 1 to 5 or 10 mg/kg body weight enhanced its therapeutic efficacy and extended its duration of action. This is explained by the short half-life of capsazepine (*Douat, Vachon & Beaudry, 2011*) reaching a maximum of 70 min with a dose of 10 mg/kg body weight. However, increasing the HC-067047 dose from 10 to 20 mg/kg rat body weight only prolonged its inhibitory effect. The three doses of ruthenium red used (0.5, 1 and 2 mg/kg body weight) were all effective for treatment of thermal hyperalgesia with different time spans. Only the 2 mg/kg reversed the mechanical hyperalgesia.

## CONCLUSION

Our novel model of perineural cancer cell invasion in Copenhagen rats successfully simulates several types of cancer-induced peripheral neuropathy and peripheral neurocarcinoma in humans. The decreased expression of TRPV1 and TRPV4 is associated with high activation. The increased expression of TLR4 and ERK1/2 reveals immune response and tumor progression, respectively, and their ultimate decrease is an indicator of nerve damage. The role of TRPV1 and TRPV4 in transducing cancer-induced hyperalgesia is investigated. The efficacy of capsazepine, HC-067047 and ruthenium red for reversal of thermal and/or mechanical hyperalgesia is highlighted.

## ABBREVIATIONS

| | |
|---|---|
| **AHF** | alcohol- and high fat |
| **ANOVA** | analysis of variance |
| **CSV** | comma-separated value |
| **DAPI** | diamidino-2-phenylindole |
| **DMSO** | dimethyl sulfoxide |
| **DRG** | dorsal root ganglia |
| **ERK1/2** | extracellular signal-regulated kinases |
| **HRP** | horseradish peroxidase |
| **LPA** | lysophosphatidic acid |
| **LPS** | lipopolysaccharide |
| **LSM** | laser scanning microscope |
| **PDB** | protein data bank |
| **PIV** | pixel intensity values |
| **RMSD** | root-mean-square deviation |
| **ROI** | region of interest |
| **S.C.** | subcutaneous |
| **SDS-PAGE** | sodium dodecyl sulphate-polyacrylamide gel electrophoresis |
| **SN** | sciatic nerve |

| **STING** | stimulator of interferon genes |
|---|---|
| **TG** | trigeminal ganglia |
| **TIF** | tagged image file |
| **TLR4** | toll-like receptor 4 |
| **TRPA1** | transient receptor potential ankyrin 1 |
| **TRPV1** | transient receptor potential vanilloid 1 |
| **TRPV4** | transient receptor potential vanilloid 4. |

## ACKNOWLEDGEMENTS

Thanks are due to our colleagues from Department of Anesthesiology and Operative Intensive Care Medicine, Charité Faculty of Medicine, Humboldt–Universität zu Berlin, Germany and Department of Biochemistry, College of Pharmacy, Al-Azhar University, Egypt who contributed to this work.

## ADDITIONAL INFORMATION AND DECLARATIONS

### Funding

This work was supported by the Egyptian Ministry of Higher Education (Cultural Affairs and Missions Sector) for two years and the German Foundation of Prof. K.H. René Koczorek (Grant #: IA89838780) for one year. The funders had no role in study design, data collection and analysis, decision to publish, or preparation of the manuscript.

### Grant Disclosures

The following grant information was disclosed by the authors:
Egyptian Ministry of Higher Education: Cultural Affairs and Missions Sector.
German Foundation of Prof. K.H. René Koczorek: IA89838780.

### Competing Interests

The authors declare that they have no competing interests.

### Author Contributions

- Ahmad Maqboul conceived and designed the experiments, performed the experiments, analyzed the data, prepared figures and/or tables, authored or reviewed drafts of the paper, approved the final draft.
- Bakheet Elsadek conceived and designed the experiments, authored or reviewed drafts of the paper, approved the final draft.

### Animal Ethics

The following information was supplied relating to ethical approvals (i.e., approving body and any reference numbers):

The study was approved by the State Office for Health and Social Affairs (LAGeSo, Berlin, Germany) and in adherence to the guidelines and the standards of Charité – University of Medicine Berlin (Project Code: G 0314/13).

## Data Availability

The following information was supplied regarding data availability:

The raw data is provided as Supplemental Dataset Files.

## Supplemental Information

Supplemental information for this article can be found online at http://dx.doi.org/10.7717/peerj.4622#supplemental-information.